\documentclass[12pt, preprint]{aastex}

\usepackage{graphicx}
\shorttitle{A broken solar type II radio burst}

\shortauthors{Kong et al.}

\usepackage{color}

\begin{document}

\title{A broken solar type II radio burst induced by a coronal shock
propagating across the streamer boundary}

\author{X. L. Kong\altaffilmark{1,2}, Y. Chen\altaffilmark{1},
G. Li\altaffilmark{3,1}, S. W. Feng\altaffilmark{1}, H. Q.
Song\altaffilmark{1}, F. Guo\altaffilmark{4},
 and F. R. Jiao\altaffilmark{1}}
\altaffiltext{1}{Shandong Provincial Key Laboratory of Optical
Astronomy and Solar-Terrestrial Environment, School of Space
Science and Physics, Shandong University at Weihai, Weihai, China
264209; yaochen@sdu.edu.cn} \altaffiltext{2}{State Key laboratory
for Space Weather, Center for Space Science and Applied Research,
Chinese Academy of Sciences, Beijing 100190,
China}\altaffiltext{3}{Department of Physics and CSPAR, University
of Alabama in Huntsville, Huntsville, AL 35899,
USA}\altaffiltext{4}{Department of Planetary Sciences and Lunar
and Planetary laboratory, University of Arizona, Tucson, AZ 85721,
USA}

\begin{abstract}
We discuss an intriguing type II radio burst that occurred on 2011
March 27. The dynamic spectrum was featured by a sudden break at
about 43 MHz on the well-observed harmonic branch. Before the
break, the spectrum drifted gradually with a mean rate of about
-0.05 MHz s$^{-1}$. Following the break, the spectrum jumped to
lower frequencies. The post-break emission lasted for about three
minutes. It consisted of an overall slow drift which appeared to
have a few fast drift sub-bands. Simultaneous  observations from
the Solar TErrestrial RElations Observatory (STEREO) and the Solar
Dynamics Observatory (SDO) were also available and are examined
for this event. We suggest that the slow-drift period before the
break was generated inside a streamer by a coronal eruption driven
shock, and the spectral break as well as the relatively wide
spectrum after the break is a consequence of the shock crossing
the streamer boundary where density drops abruptly. It is
suggested that this type of radio bursts can be taken as a unique
diagnostic tool for inferring the coronal density structure, as
well as the radio emitting source region.
\end{abstract}


\section{Introduction}

It is generally believed that type II radio bursts are caused by
electrons accelerated at magnetohydrodynamic shocks driven by solar
eruptions.  In the dynamic spectra recorded by ground-based or
space-borne radio spectrographs, type II bursts are often identified
as narrow stripes in the metric to kilometric wavelength range which
 drift gradually from high frequency to low frequency due to an
outward propagation of the electron source along with the shock
\citep{wid50,nel85}. Sometimes two stripes with a frequency ratio
about two are observed, being interpreted as the fundamental (F)
and the second harmonic (H) emissions generated via plasma
radiation mechanism at frequencies determined by the local plasma
density \citep{gin58}. The frequency drift rate of type II radio
bursts can be used to obtain the shock propagation speed by
converting the emission frequencies into coronal heights assuming
a coronal density model
\citep[e.g.,][]{rei03,vrs01,vrs02,vrs04,cho07,cho11}.

Complex and sometime elusive morphological features are frequently
present in type II dynamic spectra. For instance, the
aforementioned F and H branches may further split into two bands
causing the well-known band-splitting phenomena
\citep{sme74,sme75,vrs01}. There may exist two or more type II
bursts that occur closely in time, referred to as multi-band
events. These can be due to different shocks of different solar
drivers (i.e., flares or CMEs) or different locations along a
single shock front \citep{rob82,nel85,man04,sha05,cho11}. It has
also been known that many type II bursts exhibit intermittent
emissions rather than a continuous band. Indeed, according to
\citet{can05}, the most common type II bursts observed by
Wind/WAVES \citep{bou95} belongs to the group of ``blobs and
bands''. In addition, there are type II bursts that have
``herringbones'', which are short-lasting and rapid-drifting
emissions that extend to both higher and lower frequencies from an
overall slow-drifting backbone structure
\citep{rob59,cai87,zlo93,man05}. Apparently, these complex
spectral features are the consequences of the underlying physical
processes involved in shock formation and propagation, electron
acceleration, and the radio emitting processes. Understanding
these features can be challenging, but nevertheless are very
helpful to better decipher the underlying physical processes. We
note that simultaneous high-resolution and high-cadence solar
imaging observations such as those from the Solar TErrestrial
RElations Observatory \citep[STEREO;][]{kai08} and the Solar
Dynamics Observatory \citep[SDO;][]{pes11} can provide valuable
information on the eruption driver which can be crucial in
interpreting the observed radio bursts.

In a recent study, \citet{fen12} explored the possibility of using
type II spectral shape to infer the location of the shock-radio
emission source. They reported the presence of spectral bumps in
two solar eruption events and proposed that the spectral bumps
were caused by the shock-radio emitting region (presumably a
portion of the CME-driven shock) entering the dense streamer
structure from a less dense coronal material outside of the
streamer. In this study, we examine another metric type II event
that occurred on 2011 March 27 which showed different
morphological features as those studied in \citet{fen12}. The
paper is organized as follows. In section~\ref{sec_obser} we
describe the major features of the type II spectrum, including a
sudden drop of the frequency and the following
 intermittent fast drifting. Details of the associated solar eruption are also
presented. In section~\ref{sec_interp} we analyze the radio
dynamic spectrum and the coronal imaging data to propose a
scenario for the origin of the observed radio burst. The last
section provides our conclusions and discussion.

\section{Observations}
\label{sec_obser}

Between 00:00-01:00 UT on 2011 March 27, several ground-based
stations, including Culgoora \citep{pre94}, Learmonth, and BIRS
\citep[Bruny Island Radio Spectrometer;][]{eri97}, recorded the
type II event. In this paper, we mainly use the data of BIRS which
has a frequency range of 6-62 MHz and a time cadence of 3 seconds.
We also make use of full-disk imaging and coronagraph observations
from the STEREO and the SDO satellite. These instruments have
unprecedented high cadences and sensitivities: the
Extreme-Ultravilot Imagers (EUVI) on the Sun Earth Connection
Coronal and Heliospheric Investigation
\citep[SECCHI/STEREO;][]{how08} has a field of view (FOV)
extending
 to 1.7 $R_\odot$  and a time cadence of 2.5 minutes in 195
\AA; the FOV and time cadence of the inner coronagraph (COR1) of
SECCHI are 1.4-4 $R_\odot$ and 5 minutes; the FOV and time cadence
of the Atmospheric Imaging Assembly \citep[AIA;][]{lem11} onboard
SDO (we use the 304 \AA\ and the 171 \AA\ channels) are 0-1.5
$R_\odot$ and 12 seconds.

\subsection{A broken type II dynamic spectrum}
\label{sec_typeII}

Figure 1 displays the GOES soft X-ray flux and the associated
radio dynamic spectrum recorded by BIRS in the range 10-62 MHz and
Learmonth in the range of 62-75 MHz. Type III bursts occurred
between 00:13:30 UT and 00:16:30 UT, that were temporally
coincident with the X-ray flux due to the C3.2 flare are clearly
seen. The type II burst started about 10 minutes later when the
X-ray flux already declined to the background level. Before 00:30
UT, both the F and H branches can be identified. After this time,
the radio signals below 25 MHz became intermittent as a
consequence of ionospheric absorption and radio noises.
Nevertheless, we can still discern a few patches of emissions in
the fundamental branch. Two such patches are indicated by the
white arrows in the figure, one at 00:34 UT ($\sim$ 19 MHz) and
the other at 00:35:30 UT ($\sim$ 14 MHz). Comparing to the
fundamental branch, the harmonic branch (40-25 MHz) is more
pronounced and clear and contains more recognizable details.
Therefore we will consider only the Harmonic component of the
emission in this study.

The center of the H branch started around 60 MHz. The most
prominent feature of the H branch is the change of the spectral
slope and the presence of a breaking point separating the whole
spectrum into two parts. The break occurred at 00:33 UT and $\sim$
43 MHz. Before the break, the spectrum drifted gradually in six
minutes from 60 to 43 MHz with a mean drift rate of about -0.05
MHz s$^{-1}$. After the break, the spectrum continues as a type II
burst at lower frequencies. There was no gap in time and frequency
between the emission before the break and that after the break,
and the overall emission profiles before and after the break are
very similar although there are differences in details. Because of
these, we suggest that the whole radio spectrum corresponds to one
single underlying physical process and the sudden drop in
frequency is due to a sudden change of the background environment
(e.g. the density). However, other possibilities of interpretation
such as multiple type II bursts (e.g. Nelson \& Melrose 1985) and
two distinct type II sources by a single CME-driven shock (e.g.
Mancuso \& Raymond 2004) cannot be ruled out.
%
%
The after-break spectrum lasted for about three minutes.
The bandwidth after the
break was somewhat wider than that before the break ($>$ 10MHz).
The overall drift after the break seemed to be slower than that
before the break, but a careful examination suggests that it was
composed of a few fast-drifting bands. The first such band drifted
rapidly at the spectral break from 43 MHz to nearly 27 MHz in
about one minute with an average drift of -0.27 MHz s$^{-1}$. An
adjacent fast-drift band can be also identified from Figure~1
which has a similar frequency range and drift rate.

It should be pointed out that these fast-drifting bands are
different from the usual herringbone structures that are commonly
seen in  type II radio bursts. Indeed, in this event, herringbone
structures can be identified between 00:27 UT - 00:33 UT before
the spectral break. These structures are caused by fast electrons
moving along field lines and therefore are practically
``vertical'' structures (i.e. with a very fast frequency drift).
We are not concerned with these structures in this study.

%

The solid lines in Figure~\ref{fig1} are given by fitting the
pre-break spectrum with a two-fold Newkirk density model of the
ambient corona \citep{new61}. The shock speed is taken to be 600
km s$^{-1}$. The dashed line is an extension of the fitting to the harmonic
 branch. From the fitting, we can deduce that the type II started at $\sim$
2.0 $R_\odot$ and the spectral break was located at $\sim$ 2.3
$R_\odot$ (using the two-fold Newkirk density model). In the
following, we will compare the inferred radio emitting heights
from the fitting of the radio spectrum to those from direct
imaging observation.

\subsection{STEREO and SDO observations of the eruption}
\label{sec_eruption}

In this subsection, we discuss relevant imaging observations of
the 2011 March 27 event.

On 2011 March 27,  STEREO A and B were separated by about
176$^{\circ}$, with STEREO A $\sim$89$^{\circ}$ ahead the SDO satellite
and STEREO B $\sim$ 95$^{\circ}$ behind.
As such, these three spacecraft formed a nearly-perfect ``T'' configuration,
 allowing us to observe the eruption from both head-on and side-ways.
Before presenting the details of the eruptive processes, we first
describe the overall magnetic and morphological features relevant
to our study.

In Figure~\ref{fig2}, we show the magnetic field configurations as
obtained from the potential-field source-surface
\citep[PFSS;][]{sch69,sch03} model based on the measurements with
SDO/Helioseismic and Magnetic Imager \citep[HMI;][]{sch11} for the
Carrington Rotation (CR) 2108. The field lines are adjusted to the
three viewing angles of the spacecraft, respectively. Coronal
images from the COR1 and the EUVI instruments onboard STEREO B and
A at about 00:10 UT are also shown in Figures~\ref{fig2}(a) and
2(c).

According to Figure~\ref{fig2}(b), there existed several active
regions on the solar disk. The source of our event was NOAA active
region (AR) 11176 at S16E04 (from solarmonitor.org) as viewed from
the earth (SDO) and denoted by the yellow arrows on the EUVI data.
Clearly, the event was observed by SDO from the front as a solar
disk event and by STEREO A/B from two sides as a limb event. This
allows us to obtain simultaneously the evolutionary details of the
AR at the early stage of the eruption and its coronal responses,
which are free of projection effects that are often encountered
from observations at a later stage.

The equatorial streamer corresponds to the large-scale closed
field lines striding over ARs 11176 and 11177. The vertical black
lines atop the streamer at 2.3 $R_\odot$ in Figure~\ref{fig2}(a)
and (c) indicate the estimated location of the streamer cusp. The
dashed lines, which lie about 4 degrees above the equator,
represent the center of the streamer. The vertical black arrows on
the COR1 data point to a coronal cavity structure. The cavity
reached a height of $\sim$ 1.8 $R_\odot$ with a width estimated to
be about 0.05 $R_\odot$. Coronal cavities are usually regarded as
prominence supporting structures with relatively low density and
strong magnetic field \citep[see][]{tan74,tan95,eng89,gib06}. This
is consistent with the SDO and STEREO observations of this event.

We now examine the pre-eruption source properties and how the
eruption disturbed the surrounding plasma environment. The upper
panels of Figure~\ref{fig3} show the SDO data for the source
region. The HMI field measurements at 00:00:32 UT are shown in
Figure~\ref{fig3}(a), and the AIA 304 \AA\ image at 00:00:08 UT in
Figure~\ref{fig3}(b), both with an FOV of 0.3$\times$0.3
$R_\odot$. We can see that the source region was featured by a
S-shape bright structure as viewed in 304 \AA, which is above the
bipolar magnetic field region separated by a neutral line with
similar shape. There were many dark filament structures near the
S-shape structure, possibly related to the cavity discussed above.
Two minutes after the first observation of the eruption by AIA at
00:08:07 UT, many overlying large-scale loop structures were
observed in the 171 \AA\ wavelength as shown in
Figure~\ref{fig3}(c) at 00:10:00 UT. At this time, the large-scale
loop structures have not been affected by the eruption.

The eruption started with a sudden brightening of the S structure
which led to the outward ejection of some bright materials, as
seen in Figure~\ref{fig3}(c). The whole process can be viewed from
the corresponding online animation. It can be seen that the
ejection resulted in significant disturbances to both the
underlying and overlying loop structures. A weak yet recognizable
disturbance was observed to spread out to the surrounding corona.
It seems that some overlying arcades were opened by the ejected
materials as seen from Figure~\ref{fig3}(d) at 00:16:00 UT. In the
mean time, the eruption was well observed by the EUVI instruments
as a limb event. The corresponding observations  by STEREO B are
shown in the lower panels of Figure~\ref{fig3}.
Figure~\ref{fig3}(f) was taken when the eruption was first
observed, and Figure~\ref{fig3}(e) was taken 5 minutes earlier and
Figure~\ref{fig3}(g) 2.5 minutes later. We can see that the ejecta
moved outwards along a highly-inclined direction with an angle of
$\sim$ 60$^{\circ}$ to the radial direction, as indicated by the
white arrow in Figure~\ref{fig3}(g). In other words, the ejecta
basically propagated out along the direction of the confining
field lines in the early stage. From the last panel and the
corresponding online animation, the ejection opened the confining
arcades, and escaped from the crossing tips of the EUV rays
beneath the aforementioned equatorial streamer. The ray tips, as
marked by the ``+'' symbol, was located at $\sim$ 0.25 $R_\odot$
above the equator and $\sim$ 1.55 $R_\odot$ from the solar center.
These observations are consistent with those observed by AIA. It
should be mentioned that considerable materials falling back
towards the sun were observed after 00:30 UT as can be seen from
 the online AIA 304 \AA\ animation. We will discuss this further in
the following.

The impact of the ejecta to the outer corona was observed with the
COR1 coronagraphs.  The corresponding original image at 00:15 UT and the
running difference images from 00:20 UT to 00:40 UT obtained by
 COR1B in an FOV of 2.0$\times$2.0 $R_\odot$ are shown in the left
side of Figure~\ref{fig4}. The eruption was first observed at
00:20 UT as an outward propagating bright front sweeping through
the streamer structure. The dashed and solid lines are the same as
those plotted in Figure~\ref{fig2} representing the possible
height of the streamer cusp and the pre-disturbed streamer axis.
We can see that when first observed by COR1B the disturbance front
was located at about 1.80 $R_\odot$, and reached 2.02 $R_\odot$
and 2.24 $R_\odot$ at 00:25 UT and 00:30 UT. The front crossed the
solid line, i.e., the estimated height of the streamer cusp,
between 00:30 UT and 00:35 UT. It further propagated out and
reached 2.7 $R_\odot$ at 00:40 UT, and became diffusive and hardly
recognizable after 00:55 UT.

As viewed from the online animation, the streamer cavity,
indicated by the black arrows on the COR1 images in Figure 2,
started to get slightly larger since 00:25 UT and moved upwards
slowly. The mean speed was only about 80 km s$^{-1}$ from 00:25 UT
to 01:00 UT, much lower than that of the front. The cavity stopped
to rise at about 01:30 UT and fell back to its pre-eruption height
at about 02:30 UT. From the nearly identical locations of the
cavity and the EUVI ray tips through which the materials escaped,
we speculate that the erupted materials run into the cavity and/or
surrounding arcades and were then confined there. The fact that
the cavity did not further expand and eventually fell back
suggests that the bright front observed by the STEREO coronagraphs
represents an outward-propagating disturbance, rather than
erupting coronal material. Careful Examination of both the COR1
and COR2 images revealed that the equatorial streamer later
returned to its pre-eruption state without suffering any
disruption. There was also no apparent signatures of mass ejection
trailing the outward-propagating disturbance. This is also
consistent with the rapid weakening of the disturbance front and
the presence of considerable falling-back material as reported
previously.

The sequence of this eruption can be summarized as the following:
The initial eruption was triggered from an AR located at one foot
of the streamer structure. The eruption followed along the
highly-inclined pre-existing coronal arcades beneath the streamer
and resulted in a bright outward-moving disturbance front (likely
a shock) sweeping through the streamer structure. There was not
enough material/energy contained in the eruption and the the
ejected material were stuck within the streamer and eventually
fell back to the sun.

\section{Physical origins of the broken dynamic spectrum }
\label{sec_interp}

\begin{figure}
 \includegraphics[width=0.8\textwidth]{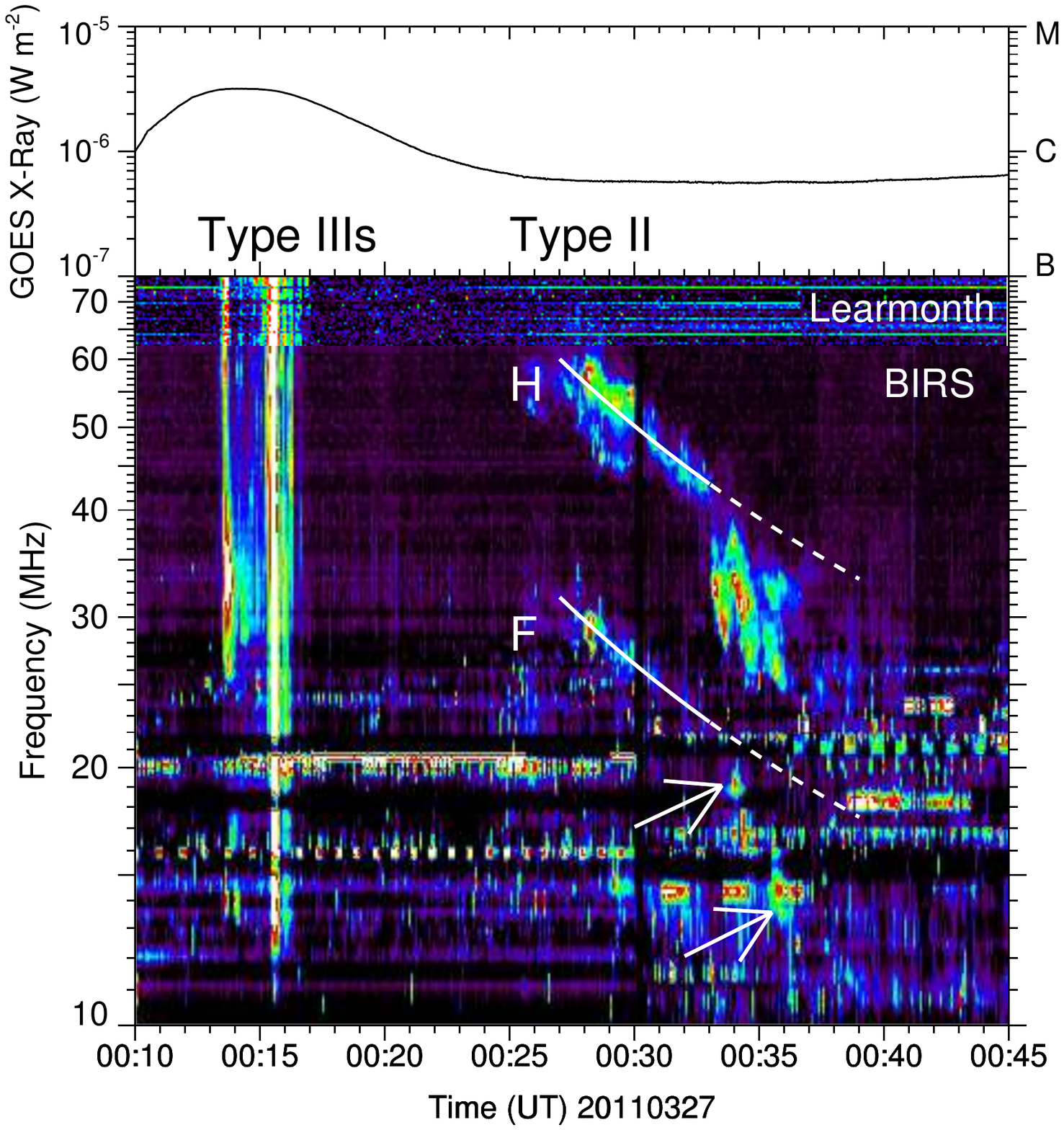}
\caption{The upper panel plots the GOES soft X-ray flux, showing a
C3.2 flare at the time of type IIIs onset. The lower panel is the
radio dynamic spectrum from BIRS (10-62 MHz) and Learmonth (62-75
MHz). ``F" and ``H" stand for the fundamental and harmonic bands
of type II burst, respectively. Solid lines are spectral fittings
using the two-fold Newkirk density model and a shock of 600 km
s$^{-1}$, the dashed line is the extension to the harmonic
fitting. The two white arrows indicate the fundamental
counterparts of the upper post-break emission.\label{fig1}}
\end{figure}

\begin{figure}
 \includegraphics[width=0.7\textwidth]{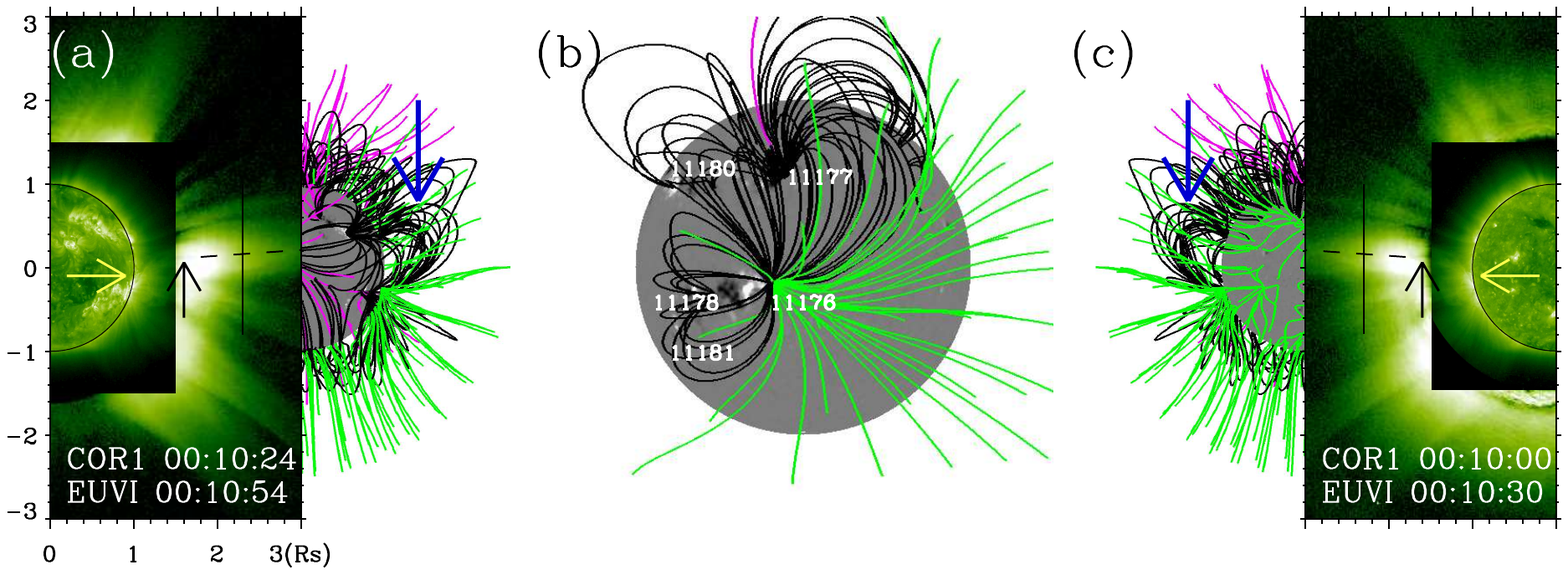}
\caption{Magnetic field configurations as obtained from the PFSS
model based on the measurements with SDO/HMI for the Carrington
Rotation (CR) 2108. The field lines are already adjusted to the
individual views of the three spacecraft. The closed lines are
colored in black, and the open inward (outward) field lines in
purple (green). Panels (a) and (c) also show coronal images from
COR1 B/A and EUVI B/A 195 \AA\  at about 00:10 UT with FOVs of
1.5$\times$3.0 $R_\odot$ and 3.0$\times$6.0 $R_\odot$. See text
for more details. \label{fig2}}
\end{figure}

\begin{figure}
 \includegraphics[width=0.7\textwidth]{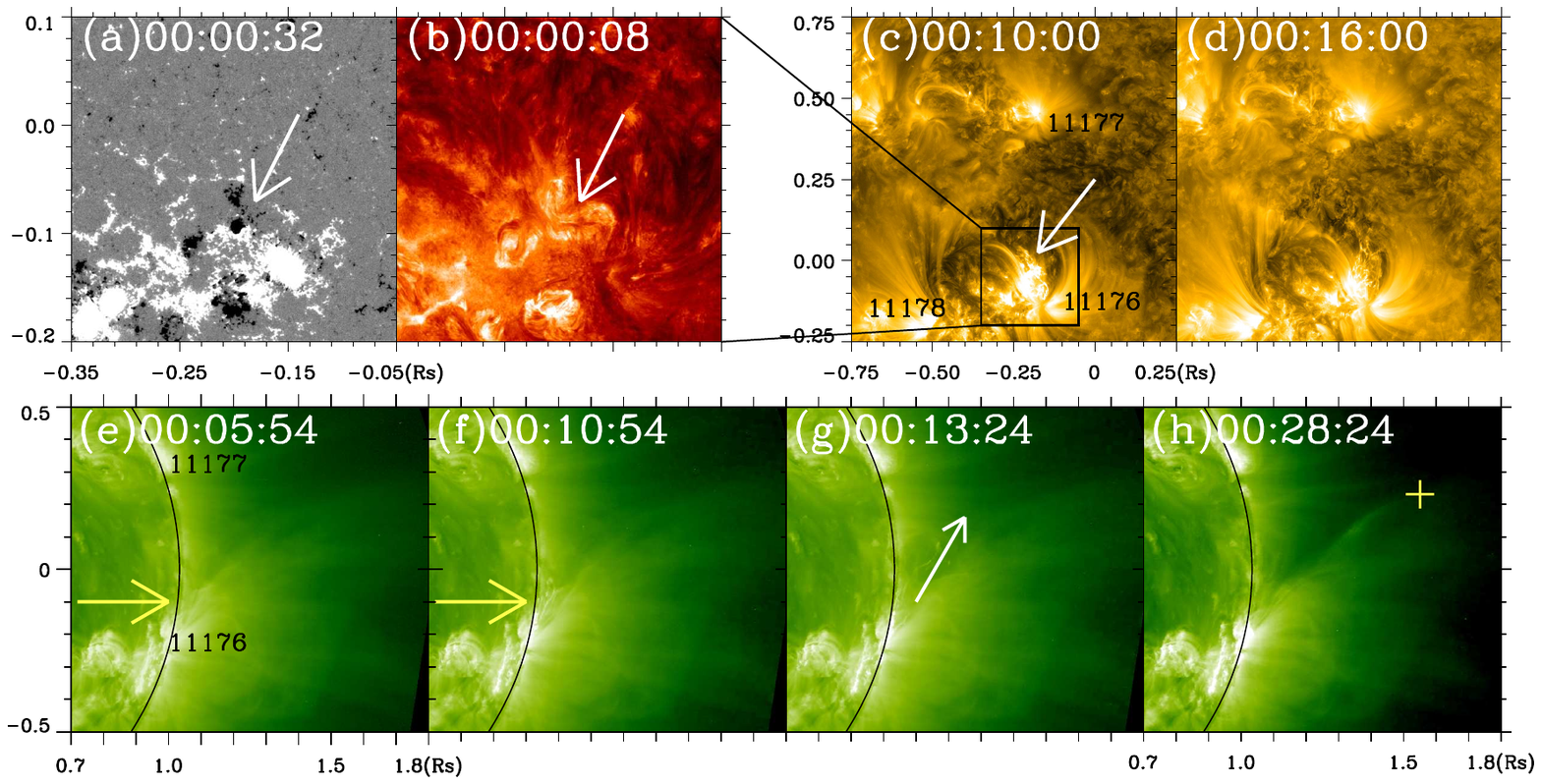}
\caption{The upper panels show the SDO data observed from the
front as a disk event. The HMI field measurements of the source
region at 00:00:32 UT are shown in panel (a), and the AIA 304 \AA\
image at 00:00:08 UT in panel (b). The location of the S-shape
bright structure are indicated by the white arrows. Panels (c) and
(d) are observed in 171 \AA, showing many overlying large scale
loop structures. The FOV for panels (c) and (d) is 1.0$\times$1.0
$R_\odot$, and 0.3$\times$0.3 $R_\odot$ for panels (a) and (b),
indicated by the black box in panel (c). The lower panels show the
EUVI B observation in 195 \AA. The yellow arrows point to the
eruption source region, i.e. AR11176, the same as in Figure 2. The
white arrow in panel (g) indicates the direction the ejecta moving
outwards in the early stage of eruption. The crossing tips of EUV
rays is marked by the plus symbol in panel (h). \label{fig3}}
\end{figure}

\begin{figure}
 \includegraphics[width=0.7\textwidth]{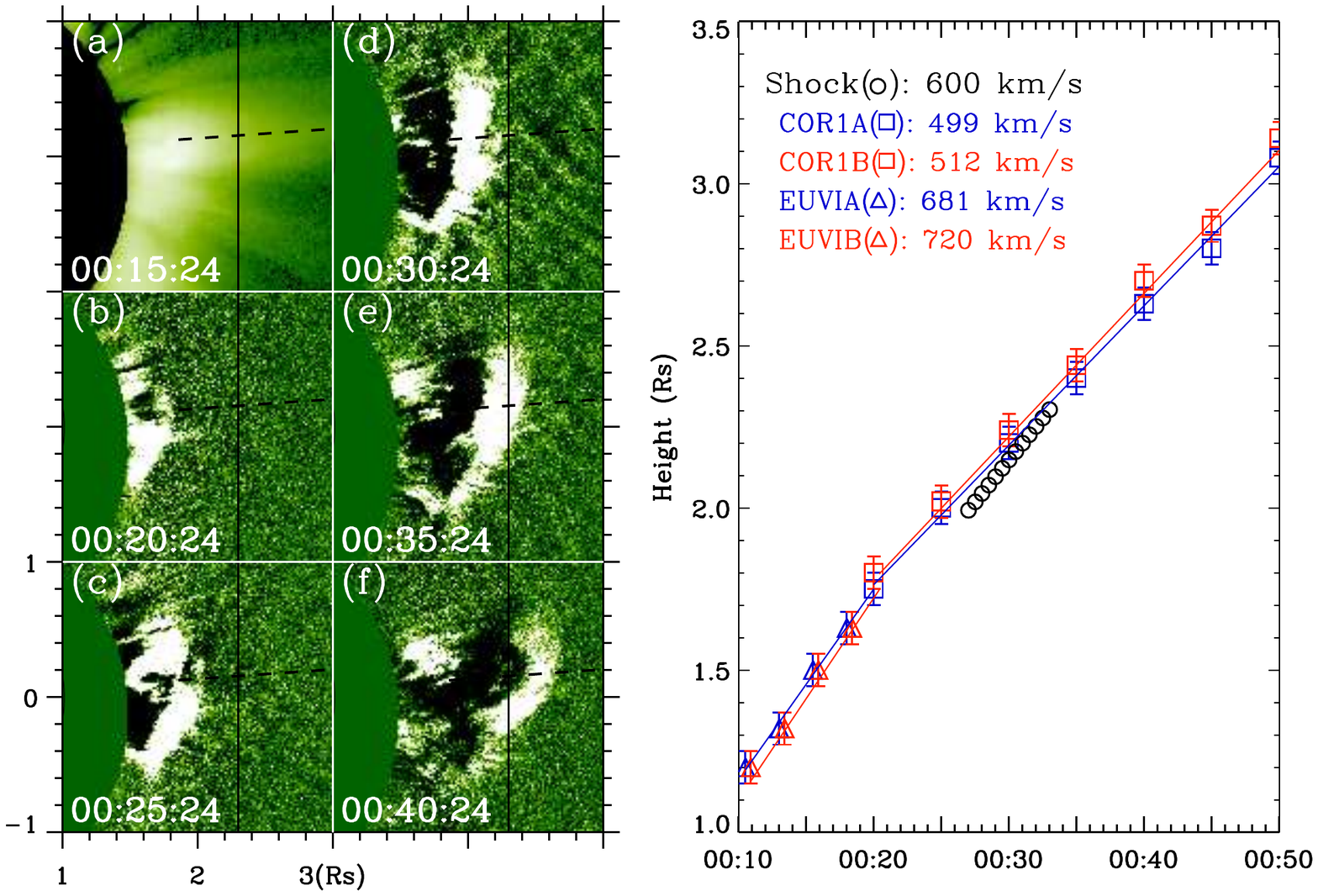}
\caption{The left side are COR1B data between 00:15-00:40UT, with
FOVs of 2.0$\times$2.0 $R_\odot$. Panel (a) is the original image
at 00:15 UT and the following are running difference images. The
black dashed lines represent the center of the streamer, and the
vertical black lines the estimated height of steamer cusp region,
the same as in Figure~\ref{fig2}. The right side plots the heights
of the disturbance front measured from STEREO, with triangles for
EUVI and squares for COR1 and in blue/red for STEREO A/B. The
error bars indicate measurement uncertainties of the heights,
estimated to be about 0.05 $R_\odot$. The lines are obtained by
linearly-fitting the EUVI and COR1 data points and the deduced
propagation speeds are shown in the upper left corner. The deduced
radio-emitting shock heights also are shown as open circles.
\label{fig4}}
\end{figure}

We now examine the physical origins accounting for the broken
feature of the type II radio burst observed in this event. To
achieve this, we first compare the heights measured from the
STEREO observations with that deduced from the dynamic spectrum
using the two-fold Newkirk density model. The fitting to the radio
data before the break was shown in Figure~\ref{fig1}, yielding a
shock speed of 600 km s$^{-1}$. The deduced radio-emitting shock
heights have been shown in the right side of Figure~\ref{fig4} as
open circles. The heights of the disturbance front measured from
the STEREO observations are also plotted as squares for COR1 and
triangles for EUVI in blue (red) for STEREO A (B). Since it is a
limb event from both STEREO A and B, measurement uncertainties are
small and estimated to be less than 0.05 $R_\odot$ (about 10
pixels). We can see that the two sets of heights agree nicely with
each other. The linearly-fitted propagation speeds of the front
are 681 and 499 km s$^{-1}$ for EUVI and COR1 of STEREO A;
 and 720 and 512 km s$^{-1}$ for STEREO B. These speeds are consistent with
that obtained by the above radio fitting method. Based on these
comparisons, we suggest that the type II burst was driven by the
shock corresponding to the bright disturbance front observed by STEREO.

We now examine the cause for the spectral break. The break
occurred at 00:33 UT between the images shown in
Figure~\ref{fig4}(d) (00:30 UT) and 4(e) (00:35 UT). During this
period, the shock front crossed the black line located at 2.3
$R_\odot$ atop the streamer. Using the two-fold Newkirk density
model,  a harmonic frequency of 43 MHz also corresponds to 2.3
$R_\odot$. Therefore it is conceivable that the spectral break was
caused by the shock crossing the streamer boundary, and the
pre-break spectrum was emitted inside the streamer. This is
consistent with the imaging observations that the event erupted at
one foot of the streamer and propagated towards the center of the
streamer.

Across the streamer boundary, the density drops faster than that
described by, e.g. the Newkirk model (which has a radial
dependence $\sim 10^{4.32/r}$). Such a larger density gradient
will lead to a faster drift of the type II radio burst.
Furthermore, since type II radio burst is expected to be sporadic
and best described as ``blobs and bands'' \citep{can05}, one
expects to find several
 such fast drifting branches, each corresponding to a radio emission region, as
the shock propagates through the streamer boundary. Indeed, this is what was observed
in the radio spectrum after the break: there seems to be three sub-branches in the
post-break radio spectrum, all with similar drift rates which are several-times
faster than that inferred from a radial density drop with a radial shock speed
of $600$ km s$^{-1}$.

\begin{figure}
 \includegraphics[width=0.8\textwidth]{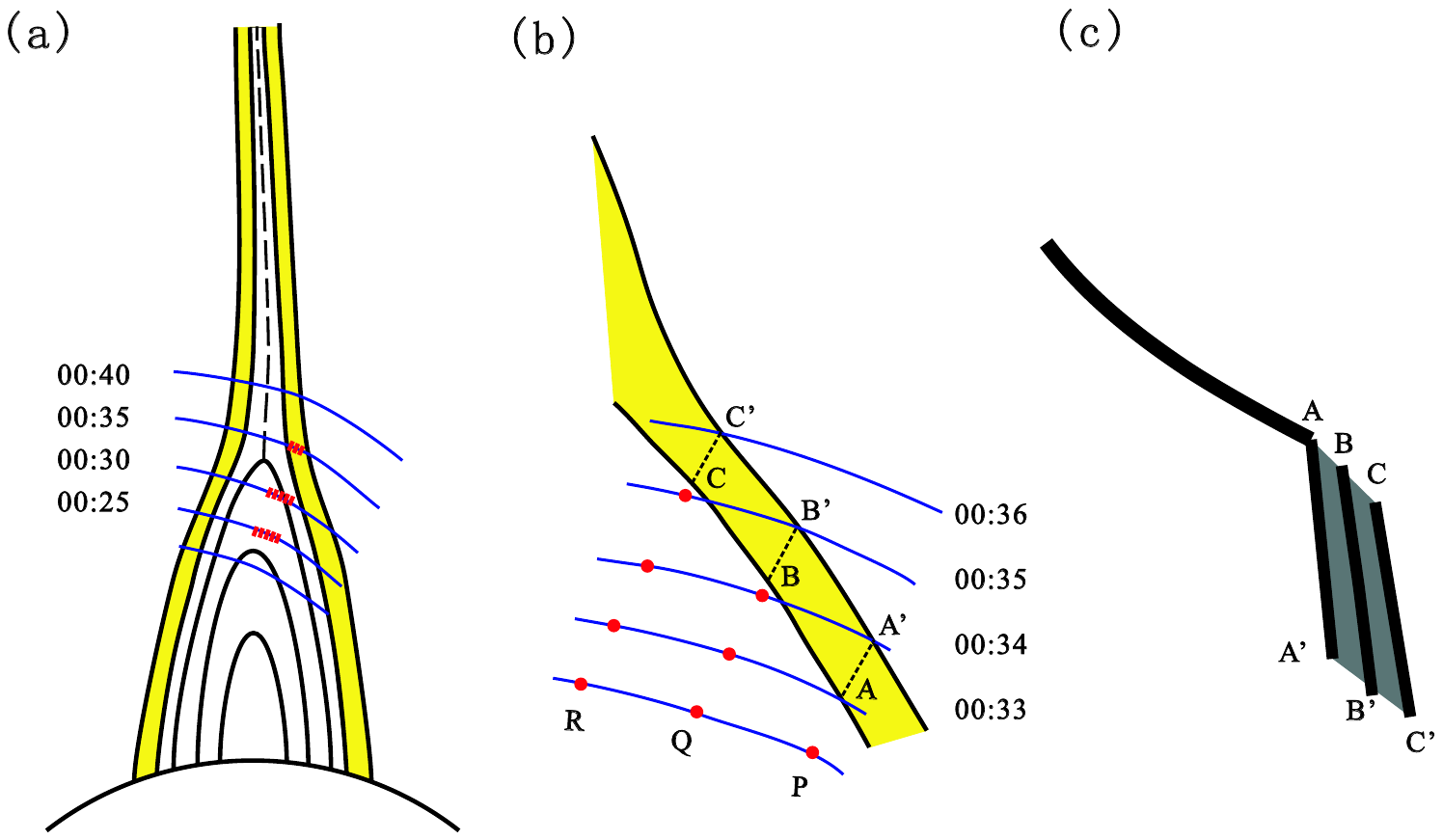}
\caption{Cartoons illustrate the physical origin of the type II
burst event. Panel (a) plots the magnetic field lines of the
streamer and locations of the shock fronts, as black and blue
curves, respectively. The streamer boundary is shaded in yellow.
The radio emitting region is indicated by the red thick segment.
Panel (b) illustrates the shock crossing the streamer boundary
after 00:33 UT and panel (c) shows the corresponding spectral
features.\label{fig5}}
\end{figure}

The process is further illustrated by the cartoons plotted in Figure~\ref{fig5}.
Figure 5(a) plots the magnetic field lines of the streamer and the locations of the
shock fronts at various times. The transition layer from inside the dense streamer to the
surrounding dilute solar wind is colored in yellow. The type II
radio emission region is indicated by the red thick segment in
Figure~\ref{fig5}(a). To illustrate the intermittent feature of
the type II radio burst, in  Figure~\ref{fig5}(b), we consider
three points P, Q, R along the shock as the major radio emission
sources. The shock front crossed the streamer cusp at a time
between 00:30 UT and 00:35 UT. The spectral break was observed at
00:33 UT corresponding to the start of the transit of the radio
source P across the streamer boundary. Before this time, all radio
sources (P, Q and R) were inside the streamer. At 00:33 UT, Point
P intersects with the transition layer at point A and later at
A$'$ (coming into the solar wind); Point Q intersects with the
transition layer at points B and later at B$'$; and Point R
intersects with the transition layer at points C and later at
C$'$. The time sequence for these intersections, as drawn in
Figure~\ref{fig5}(b) and Figure~\ref{fig5}(c), satisfies
$t(A)<t(A')< t(B)<t(B')<t(C)<t(C')$, which is consistent with
Figure~\ref{fig1}. The radiation ended soon after the radio
sources moved out of the transition layer, perhaps because the
shock weakened in the faster solar wind. Indeed, since the solar
wind is less dense and the Alfv\'en speed is expected to be
faster, the shock wave inside the streamer may ``unshock'' in the
solar wind and fail to accelerate enough electrons to excite the
type II bursts here. This is consistent with the short duration of
the post-break emission. Also note the bandwidth before the break
is governed by the density difference between P and R, which is
considerably narrower than that after the break, since the
post-break bandwidth involves a much faster density drop
introduced by the streamer boundary.

Using the measurements of the shock speed, the radio frequency
drift, and the temporal duration of the radio emission after the
break, one can in principle deduce the thickness and drop rate of
the electron density at the streamer boundary. In our event, we
can see that the emission lasts for about 2 minutes after the
spectral break, and the frequency of the H branch changes from
about 40 to 25 MHz, corresponding to a density drop from about 5
to 2 $\times$ $10^6$ cm$^{-3}$. A shock with a speed of 600 km
s$^{-1}$ propagates a distance of about 0.1 R$_\odot$ in 2
minutes. Considering the shock may not propagate perpendicularly
across the streamer boundary, this distance should be considered
as an upper limit of the radial distance of the above density
drop. Also, it should be noted that the radio emission may stop
before the shock fully cross the streamer boundary  (see Figure
5). Therefore, the values estimated above may be smaller than the
total density drop and the total thickness of the streamer
boundary. In spite of these, these values are comparable to the
electron density measurements across a coronal streamer at a
similar height of 2.33 R$_\odot$ in heliocentric distance by
\citet{str02}.

\section{Discussion and Conclusions}
In this paper we report an intriguing type II radio burst featured
by a spectral break. The radio burst is likely caused by a coronal
eruption driven shock sweeping through the streamer structure. The
spectrum before the break looks like a typical type II radio burst
with a gradually-drifting rate consistent with a shock speed of
$600$ km s$^{-1}$ moving in a corona whose density profile can be
well described by a two-fold Newkirk density model. The spectrum
after the break is composed of several fast-drifting bands, which
can be explained by shock moving across a transition layer that
has a sharper density gradient. With this interpretation, the time
of the spectral break indicates the time when the radio emission
regions (along the shock) interact with the transition layer,
therefore allowing us to use features in the type II radio bursts,
which presumably has no spatial information on the underlying
emission process, to deduce its location.

We also note that the event studied is featured by an
outward-propagating shock front which was driven by the erupting
coronal material during the early stage of the eruption. The
erupting material seemed to be confined within the overlying
streamer arcades without escaping the corona. This is inferred
from the following observations: first, the streamer, within which
the eruption took place, showed no disruption as all; second,
there were no observable signatures of mass ejection trailing the
shock front from the coronagraph data; third, a streamer cavity,
into which the ejected mass possibly ran, was disturbed yet did
not erupt during the event; and fourth, there were considerable
eruptive materials falling down towards the sun as observed in 304
\AA\ by AIA.

This study suggests that the streamer structure can play a role in
accelerating electrons and exciting type II radio bursts.
The streamer is characterized by large-scale
confining magnetic arcades in which the plasmas are basically
quasi-static without measurable outflows \citep[e.g.,][]{str02}.
Its density is also higher than the surrounding solar
wind. Therefore, the Alfv\'en speed within the streamer is likely
lower. These conditions favor the formation and strengthening of
shocks driven by a coronal mass eruption. Furthermore, a
collapsing geometry, presumably an efficient electron accelerator
\citep[e.g.,][]{zlo93,som97,mag02}, can be formed when a shock
sweeps through the closed magnetic arcades of a streamer.
To our knowledge, similar events have not been reported before and
our study is the first to examine spectral break of type II radial
bursts and explain it as an eruption within a streamer. Further
surveys to identify more events that exhibit similar spectral
features involving a coronal streamer will be carried out in the
future. Theoretical studies on the role of streamers in
shock-induced electron acceleration and type II radio bursts will
also be pursued.

\acknowledgements We are grateful to STEREO/SECCHI, SDO/AIA,
SDO/HMI and Learmonth teams for making their data available
online. We thank Dr. Stephen White and Dr. Bill Erickson for
providing the BIRS data. This work was supported by grants NSBRSF
2012CB825601, NNSFC 40825014, 40890162, 41028004, and the
Specialized Research Fund for State Key Laboratory of Space
Weather in China. H.Q. Song was also supported by NNSFC 41104113.
Gang Li's work at UAHuntsivlle was supported by NSF CAREER:
ATM-0847719 and NSF SHINE: AGS-0962658.

\newpage

\end{document}